\documentclass [12pt]{iopart}
\usepackage {iopams}
\usepackage{setstack}
\usepackage{amstext}
\usepackage{amsfonts}
\usepackage{graphicx}
\usepackage{subfigure}
\usepackage{array}
\usepackage{appendix}

\usepackage{mathrsfs}
\usepackage{color}
\usepackage{makecell}
\usepackage{adjustbox,lipsum}

\begin{document}

\title[]{Polynomial $T$-depth Quantum Solvability of Noisy Binary Linear Problem: {\em From Quantum-Sample Preparation to Main Computation}}

\author{{Wooyeong~Song}$^{1}$, {Youngrong~Lim}$^{2}$, {Kabgyun~Jeong}$^{3,2}$, {Jinhyoung~Lee}$^{4}$, {Jung~Jun~Park}$^{2}$, {M.~S.~Kim}$^{5,2}$, and {Jeongho~Bang}$^{6}$}

\address{$^1$ Center for Quantum Information, Korea Institute of Science and Technology, Seoul 02792, Korea}
\address{$^2$ School of Computational Sciences, Korea Institute for Advanced Study, Seoul 02455, Korea}
\address{$^3$ Research Institute of Mathematics, Seoul National University, Seoul 08826, Korea}
\address{$^4$ Department of Physics, Hanyang University, Seoul 04763, Korea}
\address{$^5$ QOLS, Blackett Laboratory, Imperial College London, London SW7 2AZ, United Kingdom}
\address{$^6$ Electronics and Telecommunications Research Institute, Daejeon 34129, Korea}

\vspace{10pt}

\begin{indented}
\item The first three authors (W.S., Y.L., and K.J.) contributed equally to this study and can be regarded as the main authors. 
\item Correspondence and requests for materials should be addressed to M.S.K. and J.B.
\end{indented}

\ead{\mailto{m.kim@imperial.ac.uk} and \mailto{jbang@etri.re.kr}}

\begin{abstract}
The noisy binary linear problem (NBLP) is known as a computationally hard problem, and therefore, it offers primitives for post-quantum cryptography. An efficient quantum NBLP algorithm that exhibits a polynomial quantum sample and time complexities has recently been proposed. However, the algorithm requires a large number of samples to be loaded in a highly entangled state and it is unclear whether such a precondition on the quantum speedup can be obtained efficiently. Here, we present a complete analysis of the quantum solvability of the NBLP by considering the entire algorithm process, namely from the preparation of the quantum sample to the main computation. By assuming that the algorithm runs on ``fault-tolerant'' quantum circuitry, we introduce a reasonable measure of the computational time cost. The measure is defined in terms of the overall number of $T$ gate layers, referred to as $T$-depth complexity. We show that the cost of solving the NBLP can be polynomial in the problem size, at the expense of an exponentially increasing logical qubits.
\end{abstract}

\maketitle

\newtheorem{theorem}{Theorem}
\newtheorem{lemma}{Lemma}
\newtheorem{definition}{Definition}
\newtheorem{result}{Result}
\newtheorem{estimation}{Resource Estimation (RE)}

\newcommand{\bra}[1]{\left<#1\right|}
\newcommand{\ket}[1]{\left|#1\right>}
\newcommand{\abs}[1]{\left|#1\right|}
\newcommand{\norm}[1]{\left|\!\left| #1\right|\!\right|}
\newcommand{\expt}[1]{\left<#1\right>}
\newcommand{\braket}[2]{\left<{#1}|{#2}\right>}
\newcommand{\commt}[2]{\left[{#1},{#2}\right]}
\newcommand{\round}[1]{\ensuremath{\lfloor#1\rceil}}

\newcommand{\identity}{1\!\!1}

\section{Introduction}

Owing to their simplicity, linear problems have been studied in various applications in science and engineering~\cite{Trefethen1997,Harrow2009}. However, if noise is added, it becomes exponentially difficult to solve the problems. One such challenging problem, called a noisy binary linear problem (NBLP), is defined as follows: Given a set $\mathfrak{S} = \left\{ \left( \mathbf{a}, b_\mathbf{a} \right) \right\}$ with sampled inputs $\mathbf{a}=a_0 a_1 \cdots a_{n-1} \in \{0,1\}^n$ and outputs $b_\mathbf{a} = \mathbf{a} \cdot \mathbf{s} + e_\mathbf{a} (\text{mod}~2) \in \{0,1\}$, the problem is to determine the `secret' structure of $\mathbf{s}=s_0 s_1 \cdots s_{n-1} \in \{0,1\}^n$ for all samples in the presence of noise $e_\mathbf{a} \sim B(\eta)$, where $B(\eta)$ is a Bernoulli distribution (specifically, $e_\mathbf{a}=0$ with probability $\frac{1}{2} + \eta$ and $e_\mathbf{a}=1$ with probability $\frac{1}{2}-\eta$) and $\eta \in (0, \frac{1}{2}]$. This problem is difficult to solve, and we have no better than sub-exponential sample/time complexities in classical computation~\cite{Blum2003}. This problem has thus served as a useful primitive in modern post-quantum cryptography~\cite{Regev2009}. 

Recently, Cross {\em et al.}~\cite{Cross2015} and Grilo {\em et al.}~\cite{Grilo2019} have opened the possibility that quantum computation (QC) could solve a class of NBLPs by exponentially reducing the sample/time complexities. The key feature of the proposed algorithms is the use of a quantum-superposed sample, which is defined as
\begin{eqnarray}
\ket{\psi} = \frac{1}{\sqrt{\abs{\mathfrak{R}}}}\sum_{\left( \mathbf{a}, b_\mathbf{a} \right) \in \mathfrak{R}} \ket{ \left( \mathbf{a}, b_\mathbf{a} \right) },
\label{eq:qn_samples}
\end{eqnarray}
where $\ket{ (\mathbf{a}, b_\mathbf{a}) } = \ket{\mathbf{a}} \ket{b_\mathbf{a}}$, and $\mathfrak{R} \subseteq \mathfrak{S}$ is a set of arbitrary chosen samples, and $\abs{\mathfrak{R}}$ is the cardinality of $\mathfrak{R}$. The algorithm repeatedly loads, processes, and tests the quantum sample $\ket{\psi}$ until the solution $\mathbf{s}$ is confirmed. A crucial condition for achieving a quantum speedup is that the number of samples $(\mathbf{a}, b_\mathbf{a})$ in $\ket{\psi}$ should scale exponentially with $n$; in other words, $\abs{\mathfrak{R}}$ should be $O(2^n)$. A conventional approach has hence been to employ a black-box operation (as in Eq.~(\ref{eq:qn_samples})), often called oracle, for accessing the quantum sample. However, when $\abs{\mathfrak{R}}$ is large, such an approach is not feasible because it would be costly and difficult to prepare and use a (largely-)superposed quantum sample~\cite{Aaronson2015}. In the worst case, such an approach could offset the quantum speedup achieved~\cite{Tang2021}. Therefore, although the fullest use of the quantum sample to efficiently solve the NBLP is possible in QC, it is not clear whether the hardness of the NBLP can be completely overcome. Accordingly, the security level of post-quantum cryptography has not been determined so far.  

In this paper, we present a complete analysis of the quantum solvability of the NBLP. In the analysis, we consider two essential and independent processes of the algorithm. One is loading the samples ($\in \mathfrak{S}$) into an highly entangled state $\ket{\psi}$, which is denoted by ${\cal P}_{\ket{\psi}}$. We design an optimal circuitry of ${\cal P}_{\ket{\psi}}$ by parallelising the layers of some expensive (i.e., $T$) quantum gates in the fault-tolerant level. The other process is the application of the main algorithm kernel ${\cal P}_A$, which is an optimised set of elementary gate operations. We analyse an extendable form of ${\cal P}_A$, which can cover multiple problems, and apply the result to a binary setting. The studies on ${\cal P}_{\ket{\psi}}$ and ${\cal P}_A$ have been independently performed thus far in separate contexts. For example, a recipe of optimisation of the process, similar to ${\cal P}_{\ket{\psi}}$, has been studied for a fixed architecture~\cite{Giovannetti2008-PRL}, which is designed to localize the error propagation~\cite{Matteo2020}. Likewise, the algorithms (i.e., ${\cal P}_A$ in our case) have been analysed based on a prior assumption of the quantum-sample accessibility; hence separately without any consideration of ${\cal P}_{\ket{\psi}}$. However, ${\cal P}_{\ket{\psi}}$ and ${\cal P}_A$ are systematically combined to form the quantum NBLP algorithm, and they should be studied together in a single framework\footnote{In this context, it was recently pointed out that for the discussion of the quantum solvability of noisy linear problem, not only the sample/time complexity but the superposition size of the prepared quantum-sample should be considered together~\cite{Song2022}.}. Thus, we analyse the number of repetitions of ${\cal P}_{\ket{\psi}} + {\cal P}_A$ required to determine the solution $\mathbf{s}$ in consideration of the interconnection between ${\cal P}_{\ket{\psi}}$ and ${\cal P}_A$. In the analysis, the exponential reduction in the quantum-sample complexity is derived based on a crucial condition of the solution test which has been overlooked in the previous works. This analysis allows us to account for the overall resource-consuming aspect, thereby facilitating a more comprehensive discussion on the quantum solvability of the NBLP.

The analysis is conducted in the context of the fault-tolerant QC, and we consider the $\text{Clifford}+T$ library under the assumption that an effective quantum error-correction code is embedded. We minimise the overall number of gate layers, particularly those of $T$ or $T^\dagger$ gates---which is called $T$-depth complexity~\cite{Amy2013}. Because $T$ and $T^\dagger$ are much more costly to implement than any Clifford gates in a fault-tolerant manner, the $T$-depth has often been used as a computation time performance of a quantum algorithm~\cite{Zhou2000,Fowler2009,Howard2017}. In this context, we define a computation time performance, denoted by $C$, as follows:
\begin{eqnarray}
C \equiv \left( \text{$T$-depth of ${\cal P}_{\ket{\psi}}$} + \text{$T$-depth of ${\cal P}_A$} \right) \times S,
\label{eq:cost}
\end{eqnarray}
where $S$ denotes the number of repetitions of ${\cal P}_{\ket{\psi}} + {\cal P}_A$ for the completion of the algorithm. 

We analyse the (I) {$T$-depth of ${\cal P}_{\ket{\psi}}$}, (II), {$T$-depth of ${\cal P}_A$}, and (III) repetitions $S$ and finally evaluated $C$. We note (again) that the analyses of (I), (II), and (III) are interrelated, and the quantum solvability of the NBLP cannot be described through an individual analysis of (I), (II), and (III). By managing the issues which would arise in such a comprehensive analysis (from the preparation of the quantum sample and main computation), we prove that NBLPs are polynomially solvable in the context of the $T$-depth complexity, at the expense of an exponentially increasing number of logical qubits. 


\section{Algorithm overview.} 

We briefly outline the entire procedure of the quantum NBLP algorithm.

({\bf A.1}) A state $\ket{\psi}$ of a quantum sample is prepared in the form $\ket{\psi}=\frac{1}{\sqrt{2^q}} {\sum_{\mathbf{a}}}' \ket{\mathbf{a}}\ket{b_\mathbf{a}}$, where the summation ${\sum_\mathbf{a}}'$ is of only the inputs in $\mathfrak{S}$, and $q \le n=\lceil \log_2 \abs{\mathfrak{S}} \rceil$; $q$ can be regarded as the factor that determines the size of a quantum sample $\ket{\psi}$. Here, by the term ``size of a quantum sample,'' we mean the number of the (classical) pairs $(\mathbf{a}, b_\mathbf{a})$ to be quantum-superposed in constituting $\ket{\psi}$. $\lceil x \rceil$ is the ceiling of $x$, i.e., the smallest number greater than or equal to $x$.

({\bf A.2}) Given a quantum sample $\ket{\psi}$, we run ${\cal P}_A$. Formally, ${\cal P}_A$ is given as the Bernstein--Vazirani (BV) kernel and is given by
\begin{eqnarray}
{\cal P}_A = \text{QFT}_{d}^{\otimes n+1},
\label{eq:PA_BV}
\end{eqnarray}
where $\text{QFT}_d$ is the $d$-dimensional quantum Fourier transform (QFT): $\text{QFT}_d\ket{j} = \frac{1}{\sqrt{d}}\sum_{k=0}^{d-1}\omega^{jk}\ket{k}$ with $\omega = e^{i\frac{2 \pi}{d}}$. In NBLPs, ${\cal P}_A$ becomes $\text{QFT}_{d=2}^{\otimes n+1}=\hat{H}^{\otimes n+1}$, where $\hat{H}$ is the Hadamard transform: $\ket{j} \to \frac{1}{\sqrt{2}}\sum_k (-1)^{jk}\ket{k}$ ($j,k =0,1$). The output state $\hat{H}^{\otimes n+1}\ket{\psi}$ is expressed as
\begin{eqnarray}
\frac{1}{\sqrt{2^{q+n+1}}} {\sum_{\mathbf{a}}}' \sum_{\mathbf{k}}\sum_{k^\star} (-1)^{\mathbf{a} \cdot (\mathbf{k} + \mathbf{s}k^\star) + e_\mathbf{a}k^\star} \ket{\mathbf{k}} \ket{k^\star},
\label{eq:output1}
\end{eqnarray}
where $\mathbf{k} \in \{0,1\}^n$ and $k^\star \in \{0,1\}$.

({\bf A.3}) We measure the qubit state $\ket{k^\star}$. Here, if we measure $k^\star=0$, no information on $\mathbf{s}$ can be retrieved from the remaining state, which is given by
\begin{eqnarray}
\frac{1}{\sqrt{2^{n+q}}} {\sum_{\mathbf{a}}}' \sum_{\mathbf{k}} (-1)^{\mathbf{a} \cdot \mathbf{k}} \ket{\mathbf{k}},
\label{eq:output2}
\end{eqnarray}
and the failure is returned. Otherwise (i.e., if $k^\star=1$), we obtain the remaining state
\begin{eqnarray}
\frac{1}{\sqrt{2^{n+q}}}{\sum_{\mathbf{a}}}' \sum_{\mathbf{k}} (-1)^{\mathbf{a}\cdot(\mathbf{k} + \mathbf{s}) + e_\mathbf{a}} \ket{\mathbf{k}}.
\label{eq:remain_k1}
\end{eqnarray}
By solving Eq.~(\ref{eq:remain_k1}), we obtain the candidate $\mathbf{k}$. Here, the true solution $\mathbf{s}$ can be obtained (i.e., $\mathbf{k}=\mathbf{s}$) with probability ${P(\mathbf{k}=\mathbf{s} | k^\star=1)}$, and the most exact form of the probability is $P(\mathbf{k}=\mathbf{s} | k^\star=1, \{ e_\mathbf{a} \})$. However, we drop the dependence on $\{ e_\mathbf{a} \}$ because the errors occur completely at random.

({\bf A.4}) Repeating ({\bf A.1})--({\bf A.3}), we determine the most frequently measured $\mathbf{k}$ as the true solution $\mathbf{s}$, which is referred to as ``majority voting.'' The condition of the majority voting is analysed later. Figure~\ref{fig:alg} shows a schematic of the algorithm. Additional mathematical details are provided in~\ref{appendix:1}.

\begin{figure}
\center
\includegraphics[width=0.85\textwidth]{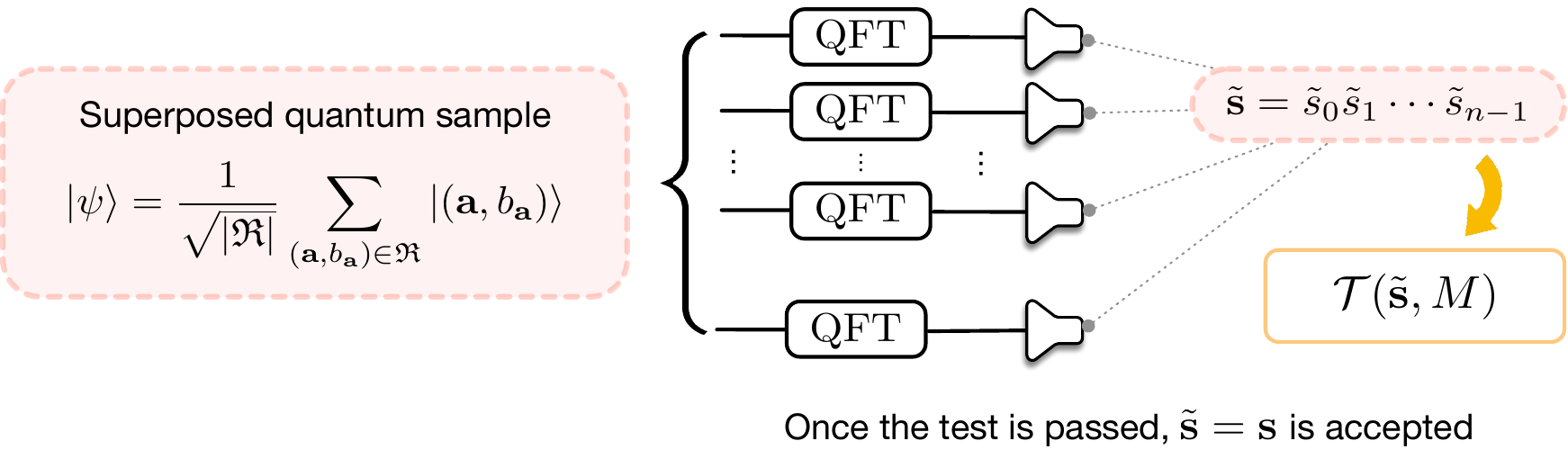}
\caption{\label{fig:alg} Schematic of the algorithm. The algorithm uses the superposed quantum sample defined in Eq.~(\ref{eq:qn_samples}) and the kernel of quantum Fourier transform (QFT). In the algorithm, a candidate fraction $\tilde{\mathbf{s}}$ is obtained and used to perform a majority voting test (for details, see the main text, or Refs.~\cite{Cross2015,Grilo2019}).}
\end{figure}

\section{Analysis (I): Resource counts for ${\cal P}_{\ket{\psi}}$.} 

Let us consider a scenario where the data, denoted by $D_{\boldsymbol\gamma}$, are addressed (or indexed) by the symbols $\boldsymbol\gamma$. The addressing (or indexing) is arbitrary and the database (or table) of $D_{\boldsymbol\gamma}$ are unsorted. We define the state of the entire data, say $\ket{T}$, as
\begin{eqnarray}
\ket{T}=\prod_{\boldsymbol\gamma \in \mathfrak{S}} \ket{D_{\boldsymbol\gamma}}.
\end{eqnarray}
Here, we note that $\ket{T}$ is not a superposition state, and each data $\ket{D_{\boldsymbol\gamma}}$ is deterministic (or equivalently, classical~\cite{Park2019}). We also emphasise that the state $\ket{T}$ itself is not computable. Our approach for analysing ${\cal P}_{\ket{\psi}}$ [or step ({\bf A.1})] is to adopt the following machinery:
\begin{eqnarray}
\left(\frac{1}{\sqrt{\abs{\mathfrak{R}}}}\sum_{\boldsymbol\gamma \in \mathfrak{R}} \ket{\boldsymbol\gamma}\right) \otimes \ket{\text{null}(D)} \otimes \ket{T} \to \left( \frac{1}{\sqrt{\abs{\mathfrak{R}}}}\sum_{\boldsymbol\gamma \in \mathfrak{R}} \ket{\boldsymbol\gamma} \otimes \ket{D_{\boldsymbol\gamma}} \right) \otimes \ket{T},
\label{eq:qdm_proc}
\end{eqnarray} 
where $\ket{\boldsymbol\gamma}$ denotes the address and $\ket{\text{null}(D)}$ is the null state in which the data brought from $\ket{T}$ are duplicated; hereafter, $\mathfrak{R}$ denotes the space of the address.

Now, we present an outline of how the machinery in Eq.~(\ref{eq:qdm_proc}) can be used to prepare $\ket{\psi}$. First, by letting $\abs{\mathfrak{R}}=2^q$, we can express the address symbol $\boldsymbol\gamma$ as a $q$-tuple of a binary number: $\boldsymbol\gamma=\gamma_0 \gamma_1 \ldots \gamma_{q-1}$, where $\gamma_j \in \{0, 1\}$ and $j=0,1,\ldots, q-1$. Subsequently, we set $\ket{D_{\boldsymbol\gamma=\mathbf{a}}}=\ket{\mathbf{a}}\ket{b_{\mathbf{a}}}$ for all samples in $\mathfrak{S}$. Such a setting is possible by matching the symbol $\boldsymbol\gamma$ is matched to the input $\mathbf{a}$. Then, from the address state $\frac{1}{\sqrt{2^q}} \sum_{\boldsymbol\gamma \in \mathfrak{R}} \ket{\boldsymbol\gamma}$, the machinery of Eq.~(\ref{eq:qdm_proc}) can provide the address-data entangled state as
\begin{eqnarray}
\ket{\Psi} = \frac{1}{\sqrt{2^q}}\sum_{\mathbf{a} \in \mathfrak{R}} \ket{\mathbf{a}} \otimes \ket{b_\mathbf{a}} \otimes \ket{T}.
\label{eq:Psi_T}
\end{eqnarray}
where the data $\ket{b_\mathbf{a}}$ are taken from $\ket{T}$ and the summation ${\sum_{\mathbf{a}}}'$ [in Eqs.~(\ref{eq:output1}), (\ref{eq:output2}), and (\ref{eq:remain_k1})] can be replaced by $\sum_{\mathbf{a} \in \mathfrak{R}}$. Lastly, we can retrieve $\ket{\psi}$ by disregarding $\ket{T}$.

\begin{figure}
\center
\includegraphics[width=1.00\textwidth]{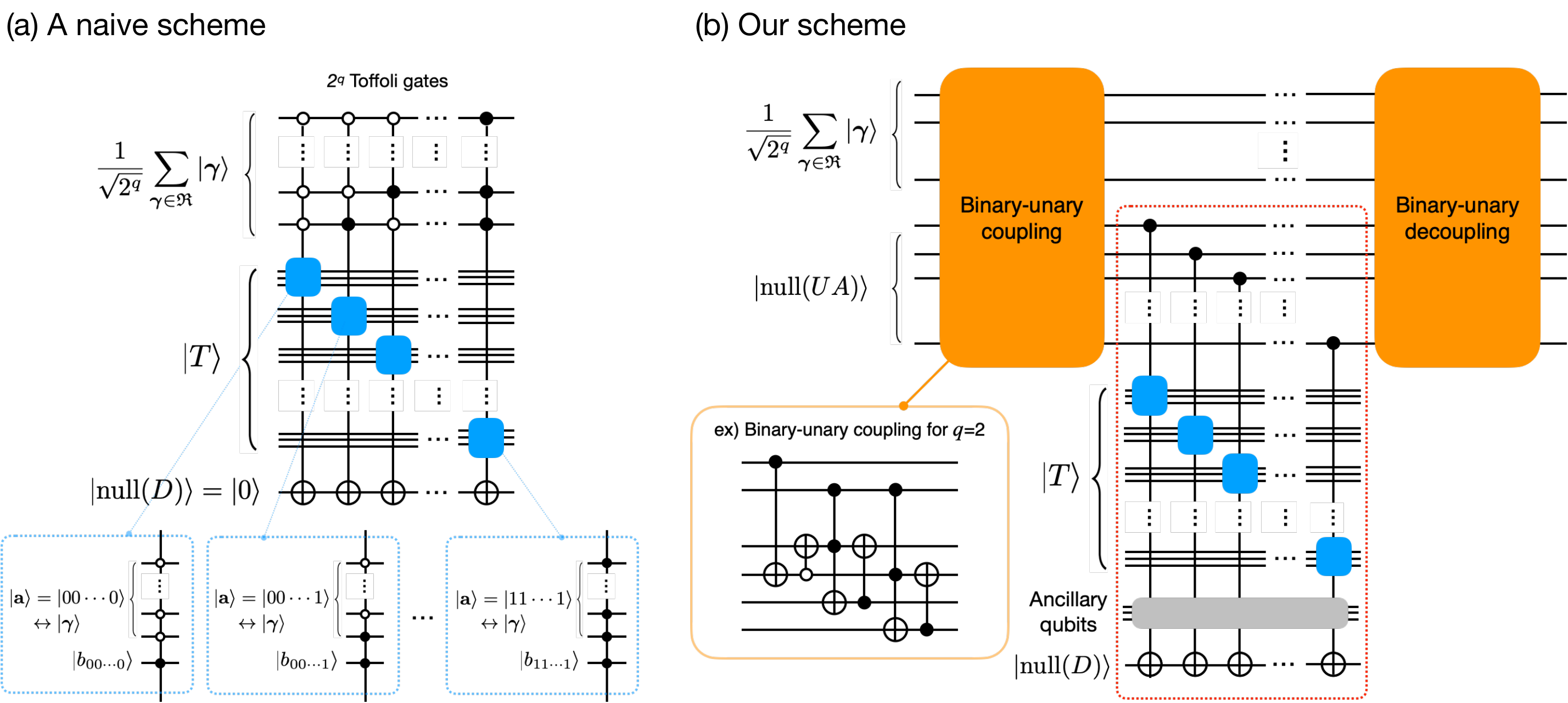}
\caption{\label{fig:basic_idea} Two schematics of the machinery in Eq.~(\ref{eq:qdm_proc}). (a) A naive approach for the data loading, where $2^q$ Toffoli gates should be implemented sequentially. In this scheme, it is impractical to reduce the $T$-depth of $O(2^q)$. (b) Our scheme designing for implementing Eq.~(\ref{eq:qdm_proc}); the unary addresses are used to bring the data. Since each unary qubit is correlated with the different data (as seen in the red dashed box), the Toffoli gates can be parallelised using ancillary qubits. Thus, if the binary-unary (de)coupling is not demanding, the advantage is straightforward (see the main text).}
\end{figure}

A naive approach for implementing the process described above is to directly load the data $\ket{b_\mathbf{a}}$ by using the Toffoli gates. However, such a data loading scheme requires an exponentially increasing $T$-depth with the address qubit size $q$, because the $2^q$ Toffoli gates should be sequentially applied (see Figure~\ref{fig:basic_idea}(a)). Therefore, our design strategy for acquiring an efficient machinery (Eq.~(\ref{eq:qdm_proc})) is to use the unary (one-hot) address encoding~\cite{Ramos2021}, as depicted in Figure~\ref{fig:basic_idea}(b). The unary bases can be written as
\begin{eqnarray}
\left\{ \ket{00\cdots01}, \ket{00\cdots10}, \ldots, \ket{01\cdots00}, \ket{10\cdots00} \right\}.
\end{eqnarray}
The unary representation does not use all available Hilbert-space, and its  advantages over the binary representation is that it simplifies the circuit structure~\cite{Paler2020}. To implement this approach, we consider two subdivided processes: 1) {\em binary-unary (de)coupling} and 2) {\em data loading}. In the subprocess 1), the unary addresses are correlated with the binary addresses. For example, for four addresses (i.e., $q=2$) one can consider
\begin{eqnarray}
&& \alpha_0\ket{00} + \alpha_1\ket{01} + \alpha_2\ket{10} + \alpha_3\ket{11} \nonumber \\
&& \to \alpha_0\ket{00}\ket{0001} + \alpha_1\ket{01}\ket{0010} + \alpha_2\ket{10}\ket{0100} + \alpha_3\ket{11}\ket{1000}.
\end{eqnarray}
The circuit for this example is presented in Figure~\ref{fig:basic_idea}(b). Subprocess 2) duplicates the data $\ket{b_\mathbf{a}}$ in $\ket{T}$ by using the unary addresses. Lastly, by decoupling the unary address qubits, we can obtain Eq.~(\ref{eq:Psi_T}). The decoupling is equivalent to the subprocess 1). Note that the unary address qubits, each of which is to be correlated with another data qubit, can easily be parallelised. Parallelisation reduces the $T$-depth complexity of the data loading considerably (as described below). Thus, if the cost of the binary-unary coupling is low, the advantage of this approach is apparent~\cite{Ramos2021}.

Let us now analyse subprocesses 1) and 2).

\begin{figure}
\center
\includegraphics[width=1.00\textwidth]{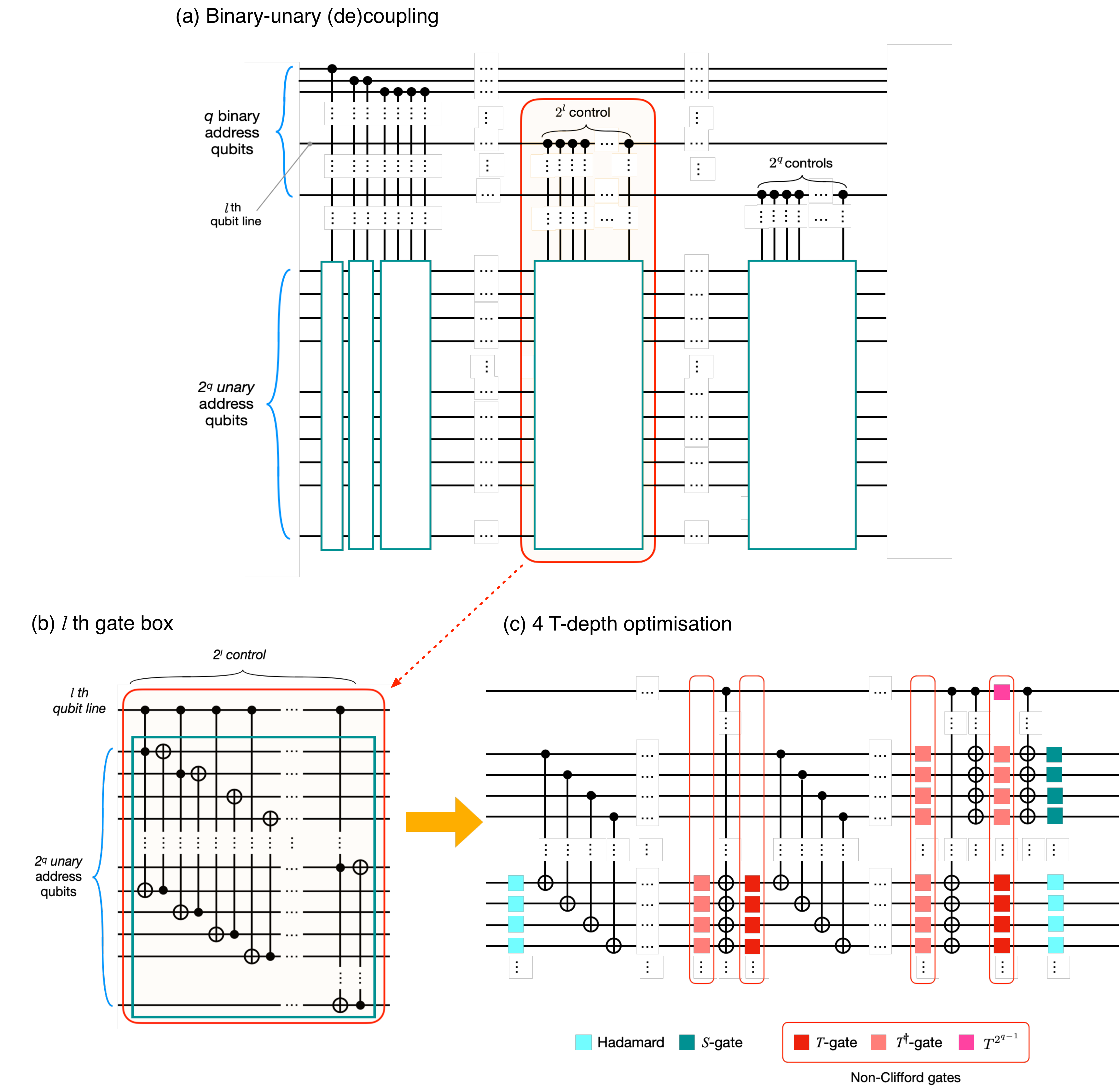}
\caption{\label{fig:binary2unary} Circuit for the binary-unary (de)coupling. (a) Generalisation of the circuit for the $q=2$ example, which comprises Toffoli and CNOT gates. (b) The generalisation part of the red box can be optimised to achieve four $T$-depths overall.}
\end{figure}

{\em 1) Binary-unary (de)coupling.}---For the analysis of subprocess 1), let us recall the circuit of the four-address example, shown in Figure~\ref{fig:basic_idea}(b). The circuit comprises Toffoli and CNOT gates. Such a circuit structure can be generalised for arbitrary $q$ address qubits, as shown in Figure~\ref{fig:binary2unary}(a), where each green box contains the gates conditioned on the $l$-th binary address qubit. The gate arrangement in the boxes can be designed generally as in Figure~\ref{fig:binary2unary}(b), where $2^l - 2$ of Toffoli gates are used in $l$-th box. It directly imposes a large $T$-depth. Thus, to minimise the depth of the circuit, we should compress the Toffoli gates~\cite{Amy2013,Selinger2013}. For this, we design an optimised circuit (termed ``four $T$-depth optimisation'') of each green box, shown in Figure~\ref{fig:binary2unary}(c), that reduces the $T$-depth of the entire process to polynomial; specifically, to $4(q-1)$.

\begin{figure}
\center
\includegraphics[width=1.00\textwidth]{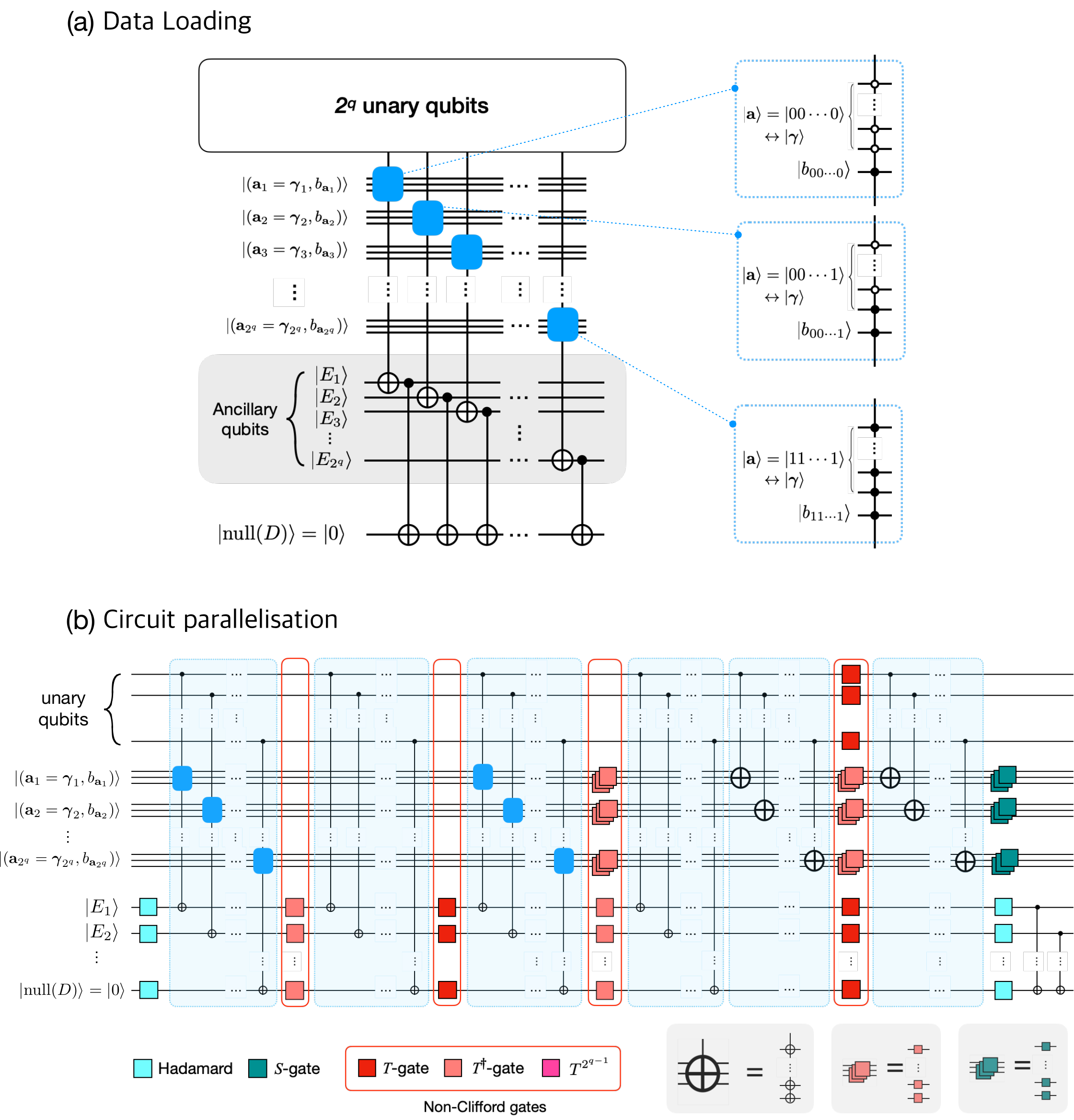}
\caption{\label{fig:data_loading} (a) Circuit for data loading; the extra ancillary qubits (denoted by $E1, E2, \cdots$) are used to avoid any constraint on the parallelisation. (b) Parallelisation of data qubits sharing Toffoli gates. Note that the gates inside the blue dashed boxes can be operated in parallel, and thus, constant $T$-depth is possible.}
\end{figure}

{\em 2) Data loading.}---In the data loading circuit, $2^q$ Toffoli gates should be used to duplicate the data $\ket{D_{\boldsymbol\gamma}}$ in $\ket{T}$ into the computable space. Thus, in the naive approach, a $T$-depth of $O(2^n)$ is required. However, since the control qubits of the Toffoli gates are each assigned one to one unary qubit in our scheme, the Toffoli gates can be implemented in parallel. This is because of the availability of the unary address. Such implementation immediately leads to the parallelisation of the $T$ gates. To avoid any restriction being imposed on the overall circuit optimisation by the control-qubit sharing of the Toffoli gates, we use the extra ancillary qubits (denoted by $E1$, $E2$, $\cdots$), as in Figure~\ref{fig:data_loading}(a). Then, every Toffoli gates can be parallelised, and the $T$-depth complexity can be optimised as $O(1)$. The detailed technique is shown in Figure~\ref{fig:data_loading}(b).

On the basis of the above analysis, our first result can be stated as
\begin{estimation}
Resource counts for implementing ${\cal P}_{\ket{\psi}}$ are as follows: The $T$-depth complexity of ${\cal P}_{\ket{\psi}}$, denoted by $T_{D,{\cal P}_{\ket{\psi}}}$, is bounded by $O(4n)$ with $q \le n=\lceil \log_2 \abs{\mathfrak{S}} \rceil$. The total number of logical qubits required to implement ${\cal P}_{\ket{\psi}}$ is determined to be
\begin{eqnarray}
\omega_{adr} + \omega_{a} + \omega_{extra} +  \omega_{D} = q + 2 \cdot 2^{q} + 1,
\label{eq:w_Ppsi}
\end{eqnarray}
where $\omega_{adr}=q$, $\omega_{a}=2^q$, $\omega_{extra}=2^{q}$, and $\omega_{D}=1$; these variables denote the number of logical qubits for the binary address, unary address, extra ancillary system and data. 
\label{est:qram_res}
\end{estimation}
 
\section{Analysis (II): Resource counts for ${\cal P}_A$.} 

Next, we consider the resource for ${\cal P}_A$. By considering the formal definition of the BV kernel [as given in Eq.~(\ref{eq:PA_BV})], we start by investigating the $T$-depth of an arbitrary $l$-qubit QFT. Usually, the quantum circuit for an $l$-qubit QFT can be synthesised with controlled-$\hat{R}_k$ gates and $\hat{H}$, where $\hat{R}_k$ denotes the single-qubit rotation and is given by $\hat{R}_k = \ket{0}\bra{0} + e^{i \pi \theta_k}\ket{1}\bra{1}$. Typically, an ideal QFT circuit requires $\frac{l(l-1)}{2}=O(l^2)$ controlled-$\hat{R}_k$ gates with $\hat{H}^{\otimes l}$, with $\theta_k = 2^{-k}~(k=1,2,\ldots,l-1)$. In practice, however, an $l$-qubit QFT can be implemented within a small fixed error $\Delta$, with $\theta_k = 2^{-k}~(k=1,2,\ldots,\beta)$ and $2 \le \beta \le l-1$. Therefore, the (so-called) approximate-QFT (AQFT) is performed using $\frac{(2l-\beta)(\beta-1)}{2}=O(l \beta)$ controlled-$\hat{R}_k$ gates. However, the condition $\beta < l-1$ implies that a finite error $\Delta$ is unavoidable because the rotation angles $\theta_k$ smaller than the threshold value $2^{-\beta}$ are discarded, limiting the choice of $\beta$. The lower bound of the order of $\beta$ is $O(\log{l})$ (Chap.~$5$ of Ref.~\cite{Nielsen2000}).

To realise an $l$-qubit AQFT circuit in a fault-tolerant manner, we can consider $\beta = O(\log{l})$. Then, all controlled-$\hat{R}_k$ gates with $\theta_k \le 2^{-O(\log{l})}$ are discarded with an error bounded by $\Delta$, and the controlled-$\hat{R}_k$ gate counts are reduced from $O(l^2)$ to $O(l \log{\frac{l}{\Delta}})$~\cite{Barenco1996}. The remaining controlled-$\hat{R}_k$ gates are decomposed into Clifford+$T$ gates, with the decomposition involving fault-tolerance overhead. Consequently, we can obtain an $l$-qubit AQFT circuit in which the number of $T$ (or $T^\dagger$) gates is $O(l \log{\frac{l}{\Delta}} \times \log{(\frac{l \log{\frac{l}{\Delta}}}{\Delta})})$, which allows the $T$-count of $O(l \log^2{l})$. For all effective QC (specifically, for $\Delta \succ l 2^{-l}$), we can neglect the dependence on $\Delta$. By noting that the $T$-depth is upper bounded by the $T$-count in general, we obtain
\begin{eqnarray}
T_{D,\text{AQFT}_{2^l}} \le T_{C,\text{AQFT}_{2^l}} = O\left(l \log^2{l}\right),
\end{eqnarray}
where $T_{C,\text{AQFT}_{2^l}}$ denotes the $T$-count of $l$-qubit AQFT. Note that in theory, $T_{C,\text{AQFT}_{2^l}}$ can be reduced more, namely from $O(l \log^2{(l)})$ to $O(l\log{l})$, by using a semi-classical AQFT~\cite{Goto2014}. Very recently, Nam {\em et al.} proposed a fully coherent AQFT that can have a $T$-count of $O(l\log{l})$~\cite{Nam2020}.

On the basis of the above analysis, we obtained the second result, which is as follows.
\begin{estimation}
We can implement ${\cal P}_A$ in the NBLP, with $T_{D,{\cal P}_A} = N/A$. The number of (logical) qubits required to execute ${\cal P}_A$ is only $O(n)$.
\label{est:alg_res}
\end{estimation}
The estimation can be validated as follows. In the the NBLP (i.e., a binary problem), ${\cal P}_A$ is the $(n+1)$-fold product of the Hadamard transform: ${\cal P}_A=\text{QFT}_{d=2}^{\otimes n+1} = \hat{H}^{\otimes n+1}$. Hence, the number of logical qubits is $n + 1$. Although the circuit may be operated with some additional ancilla qubits, $W_{{\cal P}_A}$ scales as $O(n)$. This implies zero $T$-depth complexity since controlled $\hat{R}_k$ gates are not required. Hence, {\bf RE}~\ref{est:alg_res} holds. This result is a straightforward consequence of ${\cal P}_A = \hat{H}^{\otimes n+1}$. However, an analysis of AQFT would be useful, particularly when the BV kernel is applied to a general problem setting, such as a noisy multinary linear problem.

\section{Majority-voting conditions.} 

Before analysing (III), we derive the condition for majority voting [performed in ({\bf A.4})], which has not been considered in the previous studies despite the algorithm's performance being influenced by it. First, we calculate the probability ${P_S=P(\mathbf{k} = \mathbf{s})}$ that $\mathbf{k}$ measured at ({\bf A.3}) is equal to the true solution $\mathbf{s}$. By substituting $\mathbf{k}=\mathbf{s}$ into Eq.~(\ref{eq:remain_k1}), we obtain
\begin{eqnarray}
P_S &=& P(\mathbf{k}=\mathbf{s} | k^\star = 1)P(k^\star=1) \nonumber \\
        &=& \norm{\frac{1}{\sqrt{2^{n+q+1}}} {\sum_{\mathbf{a}}}' \omega^{\mathbf{a} \cdot (2\mathbf{s}) + e_\mathbf{a}} \ket{\mathbf{s}}}^2 \nonumber \\
        &=& \frac{1}{2^{n-q+1}}\abs{\frac{1}{2^q} {\sum_{\mathbf{a}}}' (-1)^{e_\mathbf{a}}}^2,
\label{eq:Ps_cal}
\end{eqnarray}
where $P(k^\star=1)=\frac{1}{2}$. Here, we apply a useful concentration bound, the so-called Chernoff--Hoeffding (CH) inequality~\cite{Hoeffding1994}: For $t \ll O(2^{q})$,
\begin{eqnarray}
P\left( \abs{\overline{\cal U} - \mathbb{E}({\cal U}_\mathbf{a})} \ge t \right) \le 2 e^{-\frac{1}{2} 2^q t^2},
\label{eq:Hoeffding}
\end{eqnarray}
where ${\cal U}_\mathbf{a} = (-1)^{e_{\mathbf{a}}}$, $\overline{\cal U} = \frac{1}{2^q} {\sum_{\mathbf{a}}}' {\cal U}_\mathbf{a}$, and $\mathbb{E}({\cal U}_\mathbf{a})$ denotes the expectation of ${\cal U}_\mathbf{a}$. If we assume that the order of $q$ is greater than $O(\log_2{n})$, the right-hand side term in Eq.~(\ref{eq:Hoeffding}) is negligible, and $P\left( \abs{\overline{\cal U} - \mathbb{E}({\cal U}_\mathbf{a})} \ge t \right) = 0$ for a large $n$. Note that we have used the following definition [{\bf D}]: {\em If a factor is as small as $O(e^{-n})$, the factor can be negligible for a large $n$ and can be set to zero.} We then obtain the following expression: 
\begin{eqnarray}
\abs{\overline{\cal U} - \mathbb{E}({\cal U}_\mathbf{a})} < t,
\label{eq:Ps_U}
\end{eqnarray}
Using Eqs.~(\ref{eq:Ps_cal}) and (\ref{eq:Ps_U}), we can obtain the lower bound of $P_S$ such that
\begin{eqnarray}
P_S = \frac{1}{2^{n-q+1}}\abs{\overline{\cal U}}^2 > P_{S,\text{inf}}=\frac{1}{2^{n-q+1}}\abs{2\eta - t}^2,
\label{eq:lower_b_ps}
\end{eqnarray}
where we have used $\mathbb{E}({\cal U}_\mathbf{a}) = \left(\frac{1}{2} + \eta\right) - \left(\frac{1}{2} - \eta\right) = 2\eta$.

We then consider the probability ${P_F=P(\mathbf{k} \neq \mathbf{s})}$ that the measured $\mathbf{k}$ is not equal to the solution $\mathbf{s}$. For convenience, we represent $P(\mathbf{k} \neq \mathbf{s})$ as $P(\mathbf{k} = \tilde{\mathbf{s}})$, where $\tilde{\mathbf{s}} = \mathbf{s} + \boldsymbol{\phi}$. $\boldsymbol{\phi}=\phi_0 \phi_1 \cdots \phi_{n-1}$ is an arbitrary $n$-tuple of binary numbers $\phi_j \in \{0,1\}$, except for $\boldsymbol{\phi}=00\cdots0$. Then, from Eq.~(\ref{eq:remain_k1}), $P_F$ can be calculated as
\begin{eqnarray}
P_F = \frac{1}{2^{n-q+1}}\abs{\frac{1}{2^q} {\sum_{\mathbf{a}}}'  (-1)^{\mathbf{a}\cdot\boldsymbol{\phi} + e_\mathbf{a}}}^2.
\label{eq:Pf_cal}
\end{eqnarray}
Here, we recall the CH inequality in Eq.~(\ref{eq:Hoeffding}) and let ${\cal U}_\mathbf{a} = (-1)^{\mathbf{a}\cdot\boldsymbol{\phi} + e_{\mathbf{a}}}$ and $\overline{\cal U} = \frac{1}{2^q}{\sum_{\mathbf{a}}}' {\cal U}_\mathbf{a}$. It should be noted that, in this case, $\mathbb{E}({\cal U}_\mathbf{a}) = 0$ because $\mathbf{a}\cdot\boldsymbol{\phi}$ and $\mathbf{a}\cdot\boldsymbol{\phi} + e_\mathbf{a}$ are either $0$ or $1$ with probability $\frac{1}{2}$. Because $O(q)$ is greater than $O(\log_2{n})$ and $e^{-\frac{1}{2} 2^q t^2}$ is negligible by the definition [{\bf D}], we have $P\left( \abs{\overline{\cal U}} \ge t \right)=0$. Hence, we can write
\begin{eqnarray}
\abs{\overline{\cal U}} < t.
\label{eq:Pf_U}
\end{eqnarray}
By using Eqs.~(\ref{eq:Pf_cal}) and (\ref{eq:Pf_U}), the upper bound for $P_F$ is obtained as follows:
\begin{eqnarray}
P_F = \frac{1}{2^{n-q+1}}\abs{\overline{\cal U}}^2 < P_{F,\text{sup}}=\frac{1}{2^{n-q+1}}\abs{t}^2.
\label{eq:upper_b_pf}
\end{eqnarray}

We can finally specify the conditions required for the majority voting to be valid:
\begin{eqnarray}
P_{S,\text{inf}} > P_{F,\text{sup}} \Longleftrightarrow t < \eta.
\label{eq:mvote_condi}
\end{eqnarray}
If this condition is not satisfied; the possibility of a `false' solution $\tilde{\mathbf{s}}$ being identified in ({\bf A.4}) cannot be ruled out. 

\section{Analysis (III): Number of repetitions $S$.} 

Lastly, we determine the number of repetitions $S$. Let us assume that a candidate solution $\mathbf{k}$ is obtained, completing ({\bf A.1})--({\bf A.3}). The process is then repeated until $M$ candidates are collected, and finally the most frequently occurring $\mathbf{k}$ is chosen from the candidates at ({\bf A.4}). We assign $x_k = 1$ (or $x_k = 0$) when the true solution $\mathbf{s}$ (or a false solution $\tilde{\mathbf{s}}$) is measured after ({\bf A.1})--({\bf A.3}). Let $X$ be the number of times that the true solution $\mathbf{k}=\mathbf{s}$ is determined among the $M$ candidates. Then, we have $X = \sum_{k=1}^{M} x_k$ because all values of $x_k$ are independent. In such a setting, we can use a statistical inequality, namely the Chernoff bound~\cite{Mitzenmacher2017}: For any $\epsilon > 0$,
\begin{eqnarray}
P\left( \abs{X - \mu} \ge \epsilon \mu \right) \le 2 e^{-\frac{\epsilon^2}{2+\epsilon} \mu},
\label{eq:Chernoff_b}
\end{eqnarray}
where $\mu = \mathbb{E}(\identity_{\mathbf{k}=\mathbf{s}}) = M P_S$, and $\identity_{\mathbf{k}=\mathbf{s}}$ is the indicator function of $\mathbf{k}=\mathbf{s}$. By letting $2 e^{-\frac{\epsilon^2}{2+\epsilon}} \le \delta$ with $\delta \in (0,1]$, we can derive the following theorem:
\begin{eqnarray}
P\left( \abs{\overline{X} - P_S } \ge \epsilon' \right) \le \delta ~\text{iff}~ M \ge \frac{3}{\epsilon'^2}\ln{\frac{2}{\delta}},
\end{eqnarray}
where $\overline{X}=\frac{X}{M}=\frac{1}{M}\sum_{k=1}^{M} x_k$ and $\epsilon' = \epsilon P_S$ (Here, we consider a slightly weaker bound. The tight bound is given by $M \ge \frac{2+\epsilon'}{\epsilon'^2}\ln{\frac{2}{\delta}}$). This theorem implies that if we use more than ${M=\frac{3}{\epsilon'^2}\ln{\frac{2}{\delta}}}$ samples, $\overline{X}$ can be estimated within the interval $\left[ P_S - \epsilon', P_S + \epsilon' \right]$ with a probability of at least $1-\delta$. This is sometimes referred to as the sampling theorem. Since the Chernoff bound gives the minimal (Bayesian) error probability when discriminating between `a priori' and `observations', the sampling theorem translates into the following statement: Majority voting allows the identification of the true solution $\mathbf{s}$ with at least $M=\frac{3}{\epsilon'^2}\ln{\frac{2}{\delta}}$ repetitions of ({\bf A.1})--({\bf A.3}), provided the following condition is satisfied:
\begin{eqnarray}
\epsilon' < P_{S,\text{inf}} - P_{F,\text{sup}}.
\label{eq:epsilon_condi}
\end{eqnarray}
We point out that $P_{S,\text{inf}} - P_{F,\text{sup}}$ is greater than $0$ owing to the majority-voting condition in Eq.~(\ref{eq:mvote_condi}).

Furthermore, by noting that $S$ is the number of repetitions of ({\bf A.1})--({\bf A.3}), we achieve our third result, which is as follows.
\begin{estimation}
Given the constants $t$, $\epsilon$, and $\delta$, the number of repetitions $S$ is given by
\begin{eqnarray}
S=O\left(4^{n-q} \epsilon^{-2} \abs{2\eta - t}^{-4} \ln{\delta^{-1}}\right),
\label{eq:resS}
\end{eqnarray}
where we have assumed that $S = 2M$ because half of the trials of (A.1)--(A.3) will return a failure with $k^\star = 0$ (note that the factor $2$ has no influence on the order of $S$). The following crucial conditions should be satisfied:
\begin{eqnarray}
t < \eta~\text{and}~\epsilon < 1 - \frac{P_{F,\text{sup}}}{P_{S,\text{inf}}},
\end{eqnarray}
where the former is acquired from the majority-voting condition in Eq.~(\ref{eq:mvote_condi}), and the latter is derived using $\epsilon' = \epsilon P_S \ge \epsilon P_{S,\text{inf}}$ and Eq.~(\ref{eq:epsilon_condi}).
\label{est:S_res}
\end{estimation}

Note that ${\cal P}_{\ket{\psi}}$ boots up only when ${\cal P}_A$ runs with a single use of $\ket{\psi}$, and it is straightforward to determine that $S$ corresponds to the quantum-sample complexity. Accordingly, {\bf RE}~\ref{est:S_res} shows that the reduction in the complexity depends on the size of the superposition, that is, $\abs{\mathfrak{R}} = 2^q$. For example, if we use the fullest (exponential-scale) superposition of the sample with $\abs{\mathfrak{R}} = \abs{\mathfrak{S}} = 2^n$ (or equivalently, $q = n$), $S$ becomes $O(\epsilon^{-2} \abs{2\eta - t}^{-4} \ln{\delta^{-1}})$, which is consistent with the results of Ref.~\cite{Cross2015}. The opposite extreme case can also be considered, that is, using a non-superposed sample $\ket{\psi}=\ket{\mathbf{a}}\ket{b_\mathbf{a}}$ with $\abs{\mathfrak{R}}=1$ (or equivalently, $q = 0$), which still allows quantum parallelism to be processed by the BV kernel. However, in this case, $P_S$ becomes exponentially small with $n$ [ Eq.~(\ref{eq:lower_b_ps})] and is therefore negligible (based on the definition [{\bf D}]). Hence, a majority-voting condition cannot be established. Moreover, the order of $q$ is at least $O(\log_2{n})$. Note that if $q = O(\log_2{n})$, the polynomial quantum-sample complexity cannot be achieved, that is, $S = O(4^{n-\log{n}})$.

\section{Discussion}

From the results of {\bf RE}~\ref{est:qram_res},~\ref{est:alg_res}, and~\ref{est:S_res}, we can draw the following conclusion: {\em the cost $C$, defined in Eq.~(\ref{eq:cost}), can be a polynomial of the problem size $n$.} The first step to achieve the polynomial-scaling $C$ is the optimisation of the machinery of ${\cal P}_{\ket{\psi}}$ by using the unary (one-hot) sample input. Such a technique has been used to parallelise the expensive quantum gates in various contexts~\cite{Paler2020,Ramos2021}. In our case, the focus is on reducing the layers of $T$ and $T^\dagger$ gates in the context of the fault-tolerant QC. The second key enabler for our result is the BV kernel in the main computation ${\cal P}_A$, which leads to a considerable reduction in the quantum-sample complexity. However, note that the unary qubit encoding is useful for ${\cal P}_{\ket{\psi}}$, while not at all for ${\cal P}_A$. Thus, we need to transform the input from unary into binary to efficiently run ${\cal P}_A$. In summary, the polynomial $T$-depth quantum solvability of NBLPs can successfully be addressed by allowing ${\cal P}_{\ket{\psi}}$ and ${\cal P}_A$ to use favorable encodings. Note further that such a result can be achieved when the two computational features, i.e., in ${\cal P}_{\ket{\psi}}$ and ${\cal P}_A$, are analysed in a single framework. We believe that this approach will be a milestone towards confirming the overall quantum computational speedup from quantum-sample preparation to main computation.

Another insight owing to our comprehensive analysis of ${\cal P}_{\ket{\psi}} + {\cal P}_A$ is the depth-width tradeoff in the NBLP. It can be specified by Eqs.~(\ref{eq:w_Ppsi}) and (\ref{eq:resS}): roughly, $(\text{depth}) \times (\text{width})^2 \le O(4^{n})$. For example, if $q = n$, the polynomial quantum-sample complexity can be obtained (as argued in Refs.~\cite{Cross2015,Grilo2019,Song2022}). However, this suggests an exponential scale for the number of logical qubits ({\bf RE}~\ref{est:qram_res})\footnote{We note, however, that the number of logical qubits would arguably be less important than the depth of quantum circuit in terms of the algorithm speed, as the logical qubit is by definition scalable.}. By contrast, if we attempt to reduce the number of the qubits to a polynomial in $n$, for example, by letting $q = O(\log{n})$, an exponential reduction in the quantum-sample complexity cannot be achieved; and hence, the polynomial $T$-depth.

A further improvement can be achieved by developing a more efficient error-correcting code or a more efficient sample preparation scheme, which would reduce the level of noisy physical qubits.

\section*{Acknowledgements}

W.S. and J.B. thank Nana Liu for the discussions. This work was supported by the National Research Foundation of Korea (Nos. NRF-2021M3E4A1038213, NRF-2021R1I1A1A01042199, NRF-2020M3E4A1077861, NRF-2019M3E4A1079666, and NRF-2019R1A2C2005504), and the Ministry of Science, ICT and Future Planning (MSIP) by the Institute of Information and Communications Technology Planning and Evaluation grant funded by the Korean government (No.~2020-0-00890, ``Development of trusted node core and interfaces for the interoperability among QKD protocols''). W.S. acknowledges the KIST research program (2E31021). Y.L. and J.J.P. was supported by a KIAS Individual Grant (CG073301 and CG075502) at the Korea Institute for Advanced Study. M.S.K. acknowledges financial support from the Samsung GRC grant, KIAS visiting professorship, and EPSRC Quantum Computing and Simulations Hub grant.

\appendix

\section{Additional details of the algorithm}\label{appendix:1}

{\em In the absence of noise (linear function learning).}---To understand the operation of the algorithm, let us consider the case of no noise, which is often referred to as `linear function learning.' Given the sample state,
\begin{eqnarray}
\ket{\psi} = \frac{1}{\sqrt{2^q}} {\sum_{\mathbf{a}}}'  \ket{\mathbf{a}} \ket{\mathbf{a}\cdot \mathbf{s}~(\text{mod}~2)},
\label{eq:q_sample_noerr}
\end{eqnarray}
with $e_\mathbf{a}=0$ (or equivalently, $\eta = - \frac{1}{2}$), the QFTs are applied, such that
\begin{eqnarray}
\left( \underset{\text{$n$-qubit system}}{\underbrace{\text{QFT}_{d=2} \otimes \text{QFT}_{d=2}  \otimes \cdots \otimes \text{QFT}_{d=2}}} \otimes \text{QFT}_{d=2} \right) \ket{\psi}.
\end{eqnarray}
where $\text{QFT}_{d=2}$ is the Hadamard transform: $\ket{j} \to \frac{1}{\sqrt{2}}\sum_k (-1)^{jk}\ket{k}$ ($j,k = 0,1$). The output state is expressed as follows:
\begin{eqnarray}
\text{QFT}_{d=2}^{\otimes n+1} \ket{\psi} = \frac{1}{\sqrt{2^{q+n+1}}} {\sum_{\mathbf{a}}}' \sum_{\mathbf{k} \in \{0,1\}^n} \sum_{k^\star \in \{0,1\}} (-1)^{\mathbf{a} \cdot (\mathbf{k} + \mathbf{s} k^\star)} \ket{\mathbf{k}}\ket{k^\star}.
\label{eq:output_noErr}
\end{eqnarray}
Subsequently, we measured the state $\ket{k^\star}$. If $k^\star=1$ is measured using the delta function
\begin{eqnarray}
\delta_{k_j, -s_j}=\frac{1}{d} \sum_{a_j = 0}^{d-1} \omega^{a_j (k_j + s_j)},
\label{eq:delta}
\end{eqnarray}
we can achieve the final state as the true solution:
\begin{eqnarray}
\ket{\mathbf{k}}=\ket{s_0 s_1 \cdots s_{n-1}},
\end{eqnarray}
where $\omega = e^{i \frac{2\pi}{d}} = (-1)$ with $d = 2$, and the probability amplitude $\frac{1}{\sqrt{2}}$ is eliminated by the measurement of $\ket{k^\star}$. For a simpler analysis, we assume $q = n$ (hence, ${\sum_\mathbf{a}}' = \sum_{\mathbf{a} \in \{0,1\}^n}$). If $k^\star = 0$ is measured, we cannot retrieve any information of $\mathbf{s}$; that is, the algorithm returns a failure.

{\em In the presence of noise (NBLP).}---Given the sample state, that is,
\begin{eqnarray}
\ket{\psi} = \frac{1}{\sqrt{2^q}} {\sum_{\mathbf{a}}}'  \ket{\mathbf{a}} \ket{\mathbf{a}\cdot \mathbf{s} + e_\mathbf{a}~(\text{mod}~2)},
\end{eqnarray}
with non-zero noise $\eta \neq 0$, the $n + 1$ QFTs were applied as described above. We then attain the following output state:
\begin{eqnarray}
\fl \text{QFT}_{d=2}^{\otimes n+1} \ket{\psi} = \frac{1}{\sqrt{2^{q+n+1}}} {\sum_{\mathbf{a}}}' \sum_{\mathbf{k} \in \{0,1\}^n} \sum_{k^\star \in \{0,1\}} (-1)^{\mathbf{a} \cdot (\mathbf{k} + \mathbf{s} k^\star) + e_\mathbf{a} k^\star} \ket{\mathbf{k}}\ket{k^\star},
\label{eq:output_LPN}
\end{eqnarray}
which is equal to Eq.~(4) of the main manuscript. Note that we cannot use the delta function in Eq.~(\ref{eq:delta}) because unlike Eq.~(\ref{eq:output_noErr}), $\ket{\mathbf{k}}$ and $\ket{k^\star}$ are not perfectly correlated with the error term $e_{\mathbf{a}} k^\star$. Thus, Eq.~(\ref{eq:output_LPN}) allows a candidate $\mathbf{k}=\tilde{\mathbf{s}}$ that is generally not equal to the true solution $\mathbf{s}$. We can calculate the success probability, denoted by $P_S=P(\mathbf{k} = \mathbf{s})$, by substituting $\mathbf{k} = \mathbf{s} k^\star$ into Eq.~(\ref{eq:output_LPN}):
\begin{eqnarray}
P(\mathbf{k} = \mathbf{s}) &=& \frac{1}{2^{n+q}}  \norm{ {\sum_{\mathbf{a}}}'\sum_{k^\star} (-1)^{e_{\mathbf{a}}k^\star} \ket{\mathbf{s}k^\star}\ket{k^\star}}^2 \nonumber \\
    &=& \frac{1}{2^{n+q}} \sum_{k^\star} \abs{{\sum_{\mathbf{a}}}' \omega^{e_{\mathbf{a}}k^\star} }^2 \abs{\braket{\mathbf{s}k^\star}{\mathbf{s}}}^2 \abs{\braket{k^\star}{1}}^2 \nonumber \\
    &=& \frac{1}{2^{n+q+1}} \abs{{\sum_{\mathbf{a}}}' (-1)^{e_{\mathbf{a}}} }^2
\end{eqnarray}
where we use $\abs{\braket{k^\star}{1}}^2 = \frac{1}{2}$. This is equal to Eq.~(12) in the main manuscript.

\section*{References}

\bibliographystyle{iop}

\end{document}